# Size-dependent Elasticity in Materials

Nguyen Chi Huan[1] and Shailendra P. Joshi[2]

Engineering Science Programme, National University of Singapore, Singapore-117576

## 1. INTRODUCTION

The classical Euler-Bernoulli (E-B) beam is developed based on the assumption that the stress at a point is a function of the strain at that only point and the deflection is accompanied with infinitesimal strain and small rotation. However, as structural dimensions become smaller to the lattice constant of the materials, *nonlocal* elastic effects become significant and must be accounted for to render accurate predictions. Further, slender structures such as nanowires, nanotubes and so on, may experience large deflection and a nonlinear setting should be adopted. In this work, we combine the nonlocal theory of Eringen [1, 2] into the E-B beam bending together with nonlinear kinematics [3]. We briefly present the derivation and key equations of this nonlinear-nonlocal ($N_2$) beam theory and investigate the role of nonlinearity and nonlocality for simply supported nano-scaled beams.

## 2. GOVERNING EQUATIONS FOR $N_2$ E-B BEAMS

### 2.1 Nonlinear E-B beam theory:

The fundamental kinematic assumption of the E-B beam lying in the $x - z$ plane is that the displacements $(u, v, w)$, respectively in the $(x, y, z)$ directions are

$$u(x,z) = u_0 - z\frac{\partial w}{\partial x}; \qquad v = 0; \qquad w = w(x) \tag{1}$$

where $u_0$ and w are the axial and transverse displacement on the mid-plane ($z = 0$). The von Karman nonlinear in-plane strain $\varepsilon_{xx}$ at any position along the $z$ axis is

$$\varepsilon_{xx} = \frac{\partial u_0}{\partial x} + \frac{1}{2}\left(\frac{\partial w}{\partial x}\right)^2 + z\kappa = \varepsilon_{xx}^0 + z\kappa \tag{2}$$

$$\varepsilon_{xx}^0 = \frac{\partial u_0}{\partial x} + \frac{1}{2}\left(\frac{\partial w}{\partial x}\right)^2 \qquad \kappa = -\frac{\partial^2 w}{\partial x^2}$$

where $\varepsilon_{xx}^0$ is the nonlinear extensional strain and $\kappa$ is the beam curvature at its neutral axis. It is common in the E-B theory to write integral forms of stresses and the following stress resultants are used through the report

---

[1] Student
[2] Assistant Professor

$$N = \int_A \sigma_{xx} dA \quad ; \quad M = \int_A z\, \sigma_{xx} dA \tag{3}$$

where $\sigma_{xx}$ is only non-zero the normal stress, and $N$ and $M$ are the corresponding stress resultants. The principle of virtual displacement has the form

$$\delta W = \delta W_I + \delta W_E = 0 \tag{4}$$

where $\delta W_I$ and $\delta W_E$ are the virtual works due to the internal force and external forces. Explicitly,

$$\int_V \left( \rho \frac{\partial \mathbf{u}}{\partial t} \frac{\partial \boldsymbol{\delta u}}{\partial t} + \boldsymbol{f} \cdot \boldsymbol{\delta u} - \boldsymbol{\sigma} : \boldsymbol{\delta\varepsilon} \right) dV + \int_S \mathbf{t} \cdot \boldsymbol{\delta u}\, dS = 0 \tag{5}$$

where $\mathbf{u}$ and $\boldsymbol{\delta u}$ are the real and virtual displacement vectors, $\boldsymbol{\sigma}$ is the Cauchy stress tensor, $\boldsymbol{\delta\varepsilon}$ is the virtual nonlinear strain tensor, $\rho$ is the mass density of the beam, $\boldsymbol{f}$ and $\mathbf{t}$ are body force per unit volume and surface traction vectors. Substituting Eqs. (2) and (3) into Eq. (5) we obtain the following equation for E-B beams

$$\int_0^L \left[ N \left[ \frac{\partial \delta u}{\partial x} + \frac{\partial w}{\partial x} \frac{\partial \delta w}{\partial x} \right] - M \delta\kappa + f \delta u + q \delta w \right] dx = 0 \tag{6}$$

Under quasi-static conditions, we may ignore the inertial contributions and obtain the following Euler equations

$$N' + f = 0 \tag{7}$$

$$M'' + (Nw')' + q = 0 \tag{8}$$

where the number of over-primes indicates the order of partial differentiation with respect to $x$. Note that Eqs. (7) and (8) can be applied for both local and nonlocal theories alike.

**2.2 Eringen's Nonlocal theory with nonlinear kinematics:** In Eringen's linear, nonlocal theory [1], the normal stress and its stress resultants are related to the corresponding strain via

$$\sigma_{xx} - \mu \frac{\partial^2 \sigma_{xx}}{\partial x^2} = E \varepsilon_{xx} \tag{9}$$

$$N - \mu \frac{\partial^2 N}{\partial x^2} = EA\varepsilon_{xx} \qquad (10)$$

$$M - \mu \frac{\partial^2 M}{\partial x^2} = EI\kappa \qquad (11)$$

where $\mu = (e_0 L_i)^2$ is the nonlocal parameter comprising a material constant $e_0$ and an internal characteristic length $L_i$ that signifies the extent of nonlocality. Substituting Eq. (7) into Eq. (10), we obtain

$$N = -\mu f' + EA\left(u' + \frac{1}{2}(w')^2\right) \qquad (12)$$

and substituting the equation of displacement in axial direction

$$EA\left(u' + \frac{1}{2}(w')^2\right)' - \mu f' + f = 0 \qquad (13)$$

Similarly from the bending moment $M$, we obtain the equation of displacement in transverse direction

$$-\mu q'' + q - EIw'''' - \mu[Nw']''' + [Nw']' = 0 \qquad (14)$$

Equations (13) and (14) involve significant complications and we simplify them here by making a few assumptions. First, assuming $f = 0$, the terms $[Nw']'$ and $[Nw']'''$ can be expanded as

$$[Nw']' = EA\left[u'w' + \frac{1}{2}(w')^3\right]' = EA[u''w' + u'w'' + \frac{3}{2}w''(w')^2] \qquad (15)$$

$$[Nw']''' = EA[u''''w' + 3u'''w'' + 3u''w''' + w''''u' + 9w'w''w''' + 3(w'')^3 + \frac{3}{2}(w')^2 w''''] \qquad (16)$$

Further, ignoring the coupling between the derivatives of $u$ and $w$ from Eq. (15) and (16), we obtain

$$[Nw']' = \frac{3EA}{2}[w''(w')^2] \qquad (17)$$

$$[Nw']''' = EA[9w'w''w''' + 3(w'')^3 + \frac{3}{2}(w')^2 w''''] \tag{18}$$

Then, the governing equation describing the nonlinear-nonlocal ($N_2$) beam bending becomes

$$\underbrace{-\mu q''}_{nonlocal-linear} + \underbrace{q - EIw''''}_{local-linear} + \underbrace{\frac{3}{2}EA(w''(w')^2)}_{local-nonlinear} - \underbrace{\mu EA\left(9w'w''w''' + 3(w'')^3 + \frac{3}{2}(w')^2 w''''\right)}_{nonlocal-nonlinear} = 0 \tag{19}$$

## 3. Numerical Results and Discussion

The governing equation (Eq. 19) derived above can be solved numerically in order to investigate the behavior of E-B beams under various displacement and load boundary conditions. In this work, we consider a simply supported beam under a uniformly distributed load $q$. The kinematic b.c.'s at the supports are as follows

$$w(x=0) = w(x=l) = 0 \tag{20}$$

$$w''(x=0) = w''(x=l) = \frac{\mu}{EI}\left(q + \frac{3}{2}EA[w''(w')^2]\right) \tag{21}$$

Note that Eq. (21) involves a highly nonlinear b.c. We choose to simplify this by ignoring the second term in the bracket on the right hand side of the equation.

The material and geometric parameters chosen for computational purposes are

$$E = 1000 \text{GPa}; \quad L_i = 1 \text{ (nm)}; \quad L = 15 \text{ (nm)}; \quad \text{Aspect ratio} = 10;$$

These parameters are consistent with those typically adopted in the analysis of carbon nanotubes. For comparison, we analyze four cases: (a) Local-linear ($L_2$), (b) nonlocal-linear ($N-L$), (c) local-nonlinear ($L-N$) and (d) Nonlocal-nonlinear ($N_2$).[3]

Figure 1 shows the relationship between the normalized maximum deflection $w_{max}$ and normalized applied loading for a fixed $\mu$ for the four cases. As expected, the linear theories ($L_2$ and $N-L$) predict linearly increasing $w_{max}$. The nonlocality tends to increase the beam deflection with increasing $q$. In other words, in the presence of the nonlocal effect the beam appears to be elastically softer in bending for the same $q$. If the nonlinear effect is retained the beam deflections are much smaller than their linear counterparts, irrespective of whether or not the nonlocal behavior is accounted for. This is because the presence of nonlinear strain term provides an additional stiffening mechanism against deflection. Interestingly, even in the presence of the nonlinearity the nonlocal effect gives a higher $w_{max}$ for a given $q$ compared to the local theory. These basic results bring to the fore a couple of interesting aspects. Firstly, as shown in the graph, the difference between nonlocal theories and local theories is significant within the context of the internal length-scale. This difference cannot be ignored when one has to deal with nanostructures. Secondly, the effect of nonlocal theory is to increase the overall deflection, whereas the nonlinear theory tends to decrease it. This latter aspect creates a competition between the stiffening and softening mechanisms in elastic beam bending.

---

[3] We solved the system of equations using MAPLE®.

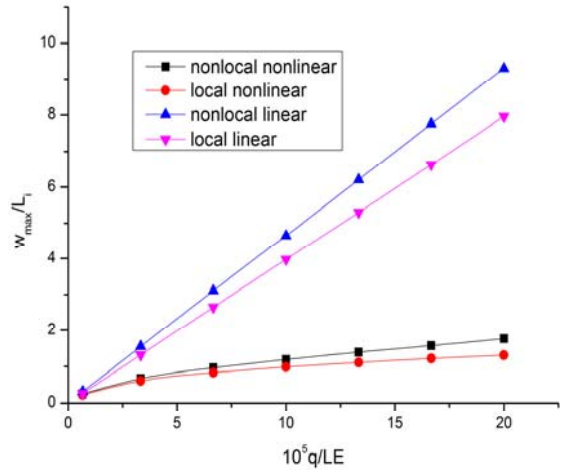

**Figure 1** Plot of the normalized maximum deflection versus normalized applied load ($\mu = 1\ nm^2$).

To the best of our knowledge, there are no results available that include both nonlinear and nonlocal effects in beam bending. However, we have compared the degenerate N-L case (Fig. 1, blue curve) with those provided by [4] and they match exactly.

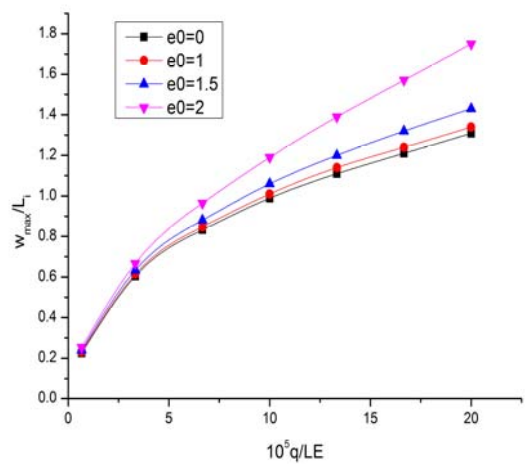

**Figure 2** Effect of material constant $e_0$ on the maximum deflection for the $N_2$ case.

As shown in Fig. 2, increasing $e_0$ (higher nonlocality) increases the deflection. Therefore, we predict that at some combination of $e_0$ and $q$ the competition between nonlinearity and nonlocality may reach saturation, beyond which one effect will dominate the other. In other words, at higher nonlocality the nonlinear effect may diminish and vice-versa. Figure 3 proves such a behavior in that a larger nonlocal parameter is chosen for the problem. For these values, the $N_2$ result nearly mimics the $L_2$ result indicating that the nonlocal and nonlinear effects tend to largely compensate each other.

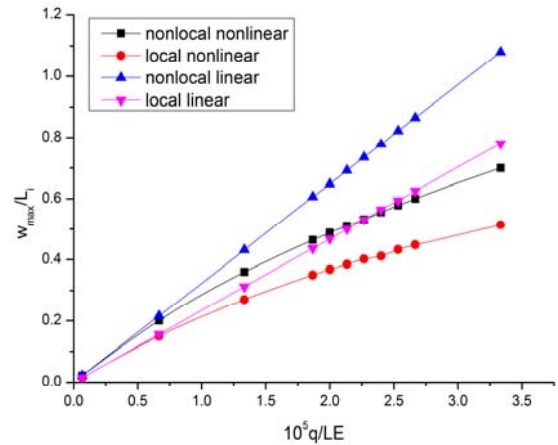

**Figure 3** Plot of the normalized maximum deflection versus normalized applied load ($\mu = 9\ nm^2$).

## 4. CONCLUSIONS

In conclusion, we have investigated the coupled nonlinear-nonlocal elastic behavior of Euler-Bernoulli beams. The governing equation is simplified by ignoring some of the complicated coupling terms. Although, this introduces approximations, it renders a useful way of investigations without significant numerical difficulties. The roles of ignored terms should be assessed. Such a formulation can also be extended to nonlinear vibration by accounting for inertia. For the particular problem of a simply supported beam subjected to uniform loading, a significant result is that the nonlocal and nonlinear effects oppose each other and modulate the overall behavior. Nonlocality assists beam deflection, but nonlinearity resists it. At critical values of nonlocal parameters the nonlinear behavior may be largely compensated by nonlocal effects.

## 5. SELECTED REFERENCES